%% file: thesistemplate.tex
\documentclass[a4paper,12pt]{scrreprt}

\usepackage{amsfonts,amsmath,amssymb,bm} 
\usepackage{booktabs} 
\usepackage{color} 
\usepackage{fancyhdr} 
\usepackage{graphicx} 
\usepackage{units} 
\usepackage{xspace} 
\usepackage{setspace}
\DeclareMathOperator*{\argmax}{argmax} 
\DeclareMathOperator*{\argmin}{argmin} 
\usepackage{array} 
\usepackage{lipsum} 
\usepackage{subcaption} 
\usepackage{rotating} 
\usepackage{todonotes} 
\usepackage[linesnumbered,ruled,vlined]{algorithm2e}

\usepackage[bookmarks=true,bookmarksopen=true,bookmarksopenlevel=2]{hyperref} 
\hypersetup{
  pdfpagemode=UseOutlines,
  pdfstartview=Fit,
  pdftitle={},
  pdfsubject={},
  pdfauthor={},
  pdfkeywords={},
  pdfcreator={pdflatex}
}

\numberwithin{equation}{chapter}
\numberwithin{figure}{chapter}
\numberwithin{table}{chapter}

\newcolumntype{V}[1]{>{\raggedright\hspace{0pt}}p{#1}} 
\newcommand{\Pics}{./PICTURES} 

\begin{document}

\setcounter{page}{-1}  
\include{TEXs/titlepage}

\cleardoublepage 

\pagenumbering{Roman} 
\setcounter{page}{1}  

\pagestyle{fancy}
\renewcommand{\chaptermark}[1]{\markboth{#1}{}}
\renewcommand{\sectionmark}[1]{\markright{\thesection\ #1}}
\fancyfoot{}
\fancyhead{}
\fancyhead[LE,RO]{\small\bfseries \thepage}
\fancyhead[LO]{\small\bfseries \rightmark}
\renewcommand{\headrulewidth}{0.5pt} 

\include{TEXs/abstractenglish}

\cleardoublepage 

\setcounter{tocdepth}{1} 
\tableofcontents
\pdfbookmark[0]{Table of Contents}{Table of Contents}
\listoffigures
\listoftables

\newpage
\pagenumbering{arabic} 

\include{TEXs/introduction}

\include{TEXs/BVF}

\include{TEXs/PCE}

\include{TEXs/Application}


\pdfbookmark[0]{Bibliography}{Bibliography}
\bibliographystyle{LITERATURE/lf_gb_ila_vt}
\bibliography{LITERATURE/literature}

\begin{appendix}
\end{appendix}

\end{document}

%% file: TEXs/titlepage.tex
\thispagestyle{empty} 

\vspace*{-2truecm}

\newlength{\HeadWidth}
\setlength{\HeadWidth}{\textwidth}

\newlength{\LogoWidthL}
\settowidth{\LogoWidthL}{\includegraphics[height=20mm]{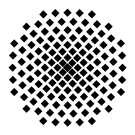}}
\addtolength{\HeadWidth}{-\LogoWidthL}
\addtolength{\HeadWidth}{-3mm} 

\newlength{\LogoWidthR}
\settowidth{\LogoWidthR}{\includegraphics[height=23mm]{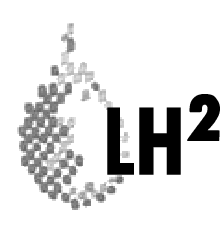}}
\addtolength{\HeadWidth}{-\LogoWidthR}
\addtolength{\HeadWidth}{-3mm} 

\newlength{\BoxBreite}
\setlength{\BoxBreite}{\textwidth}

\vspace*{-1.5cm} 
{
  \parbox[b]{\BoxBreite}
  {                               
    \includegraphics[height=20mm]{\Pics/Logo_UniS-zentriert}
    \parbox[b][22mm][c]{\HeadWidth}
    {                                       
      \fontsize{10}{11pt}\selectfont \sffamily        
      \begin{center}          
        Universit\"at Stuttgart  -  Institut f\"ur Wasser- und Umweltsystemmodellierung\\    
        {\bfseries 
          Lehrstuhl f\"ur Hydromechanik und Hydrosystemmodellierung}\\
        Prof.~Dr.-Ing.~Rainer Helmig    
      \end{center} 
      }
    \includegraphics[height=23mm]{\Pics/Logo_LH2_sw} \\ [-3mm] 
    }
  }       
\vspace{3cm}

\begin{center}
  {\bfseries {\LARGE Development and Realization of Validation Benchmarks\\[0.3cm]}}
\end{center}

\vspace{7cm}

\begin{center}
  {\large Adapted from the original Milestone Report}\\[0.5cm]
  {\normalsize Submitted by}\\[0.1cm]
  {\large Farid Mohammadi, M.Sc.}\\[0.1cm]
\end{center}

\vspace{2cm}

\begin{center}
  {\large Stuttgart, 10.06.2021}
\end{center}

\vspace{1cm}

\begin{center}
  {\large 
  \begin{tabular}{l}
  Supervisor: apl. Prof. Dr.rer.nat Bernd Flemisch\\
              PD Dr.-Ing Sergey Oladyshkin \\
              Prof. Dr. Bruno Sudret
  \end{tabular}
  }\\[0.1cm]
\end{center}

%% file: TEXs/abstractenglish.tex
\chapter*{Abstract}
\thispagestyle{empty}

In the field of modeling, the word \textit{validation} refers to simple comparisons between model outputs and experimental data. Usually, this comparison constitutes plotting the model results against data on the same axes to provide a visual assessment of agreement or lack thereof.
While comparisons between model and data are at the heart of any validation procedure, there are a number of concerns with such naive comparisons.
First, these comparisons tend to provide qualitative rather than quantitative assessments and are clearly insufficient as a basis for making decisions regarding model validity. Second, naive comparisons often disregard or only partly account for existing uncertainties in the experimental observations or the model input parameters. Third, such comparisons can not reveal whether the model is appropriate for the intended purposes, as they mainly focus on the agreement in the observable quantities.

These pitfalls give rise to the need for an uncertainty-aware framework that includes a validation metric. This metric shall provide a measure for comparison of the system response quantities of an experiment with the ones from a computational model, while accounting for uncertainties in both, in a rigorous way. To address this need, we have developed a statistical framework that incorporates a probabilistic modeling technique using a fully Bayesian approach. A Bayesian perspective on a validation task yields an optimal bias-variance trade-off against the experimental data and provide an integrative metric for model validation that incorporates parameter and conceptual uncertainty. Additionally, to accelerate the validation process for  computationally demanding flow and transport models in porous media, the framework is equipped with a model reduction technique, namely Bayesian Sparse Polynomial Chaos Expansion. 
We demonstrate the capabilities of the aforementioned Bayesian validation framework by applying it to an application for validation as well as uncertainty quantification of fluid flow in fractured porous media.

%% file: TEXs/introduction.tex
\chapter{Introduction}

Over the last century, the computational modeling in the field of porous media has witnessed tremendous improvement. After decades of development, the state-of-the-art simulators are now capable of solving coupled partial differential equations governing the complex subsurface multiphase flow system within a practically large spatial and temporal domain. 
Given the importance of computational modeling, assessment of the reliability of models in light of the purpose of a given simulation is of paramount importance. For this reliability assessment, there exists various complementary measures, such as unit testing of individual components of the computational model, comparisons with analytical solutions or other computational models. 

While assessing the simulation results with aforementioned measures could hint us towards the correctness of the model, measuring the computational model performance against experimental data can enhance the confidence in the models.
However, comparisons with experimental data can be very cumbersome, since the experiment in question has probably not been designed to meet the objective of the validation of a computational model. Moreover, disagreement with experimental data can have other reasons than the deficiency of the simulation code, such as an insufficient description of the experimental set-up or considerable uncertainties associated with the measurements. On the other hand, pure code intercomparison studies cannot ensure that a successfully participating model indeed maps the reality.

The goal of this PhD project is to formulate and conduct benchmarks, which assist in the uncertainty-aware validation of several computational models developed within the Collaborative Research Center 1313. With the intended validation benchmarks developed in this project, we plan to overcome the forgoing shortcomings. 

\section{Validation of Physical Models in the Presence of Uncertainty}

In the field of modeling, it is fairly common to use the word validation to refer to simple comparisons between model outputs and experimental data. Usually this comparison constitutes plotting the model results against data on the same axes to provide visual assessment of agreement or lack thereof.
According to Moser et al. \cite{moser2015validation}, while comparisons between model and data are at the heart of any validation procedure, there are a number of concerns with such naive comparisons.
First, these comparisons tend to provide qualitative rather than quantitative assessments. Such qualitative assessments are often essential  and informative. However, they are clearly insufficient as a basis for making decisions regarding model validity. Second, naive comparisons often disregard or only partly account for existing uncertainties in the experimental observations or the model input parameters. Without accounting for these uncertainties, it is not possible to appropriately determine whether the model and data agree. Third, such comparisons can not reveal whether the model is appropriate for the intended purposes, as they mainly focus on the agreement in the observable quantities.

These pitfalls of straightforward but naive comparisons gives rise to the need for a framework that includes a validation metric. This metric must measure the system response quantities of an experiment with the ones from a computational model, while accounting for uncertainties in both in a rigorous way. This fact is widely recognized, particularly in the statistics community, and there are a number of possible approaches, but here we employ a statistical framework.
We have developed such framework by means of a Bayesian validation framework that incorporates parameter and conceptual uncertainty. Incorporation of a fully Bayesian approach yields an optimal bias-variance trade-off against the experimental data and provide an integrative quantity for model validation. Additionally, in order to guarantee the feasibility of the Bayesian validation framework for computationally expensive models, we accelerate the computations for expensive models via model reduction techniques.

The remaining of this report is structured as follows: Chapter 2 introduces components of the Bayesian validation framework. This is followed by Chapter 3, which elucidates the contribution of a surrogate modeling to offset the computational cost. Lastly, the application of the Bayesian validation model to flow simulation models for fractured porous media is presented.

%% file: TEXs/BVF.tex
\chapter{Bayeisan Validation Framework}
\thispagestyle{empty}

\section{Validation from Bayesian Perspective}
Bayesian epistemology offers a robust framework for characterizing scientific inference since its simple concept lies in the fact that rational belief comes in degrees that can be measured in terms of probabilities. Moreover, Bayesian epistemology has resulted in the useful elucidation of notions such as confirmation. Thus, it is proven to form a viable method for data-driven validation of computer simulations and can provide a solid basis for a sound evaluation of computer simulations.

Bayesian epistemology dates back to ideas by Thomas Bayes and was reinforced by many scientists in the twentieth century, such as Bruno de Finetti, Frank P. Ramsey, and Leonard Savage, among others.
Bayesianist believes that trust comes in degrees that are measured in terms of probabilities, or probability densities when dealing with characteristics with a continuous range of values. They generally tend to proceed with the following three steps. First, they formulate plausible hypotheses related to a simulation. Second, they consider rational degrees of belief in these hypotheses. Thirdly, they apply the bayesian principles. Following these steps yields posterior probabilities that reveal the degree of trust, one shall rationally invest in the hypotheses.

What are the relevant hypotheses regarding the validation of a simulation? These could be building up or losing trust in simulations. In terms of validation, the hypothesis is whether the model is able to satisfactorily represent the real system of interest. Moreover, for assessment of physical phenomena in question, several representations, i.e. models, might exist with different approaches and assumptions to analyze the occurring processes. In this case, the hypothesis is which model within the pool of available models can represent the reality, i.e. observed values in the experiments.  
But, How can we arrive at the aforementioned probabilities? This can be achieved by updating the prior belief on the basis of Bayesian notions. For this purpose, simulation results need to be compared with data.

\subsection{Bayes' theorem}
The Bayesian approach to validation extensively exploits Bayes’ theorem. This theorem is a combination of traditional probabilities and statistics. Let us assume $A$ and $B$ are two events. The conditional probability of event $A$ given that we know that the event $B$ has occurred, $P(A|B)$, can be cast as:
\begin{equation}
\label{eq:CondProbab}
    P(A|B) = \frac{P(A \cap B)}{P(B)},
\end{equation}
where $P(B)$ denotes the probability of event $B$, which poses positive value, and $P(A \cap B)$ signifies the probability that both $A$ and $B$ occurred. Given that $P(A \cap B)$ is equal to $P(B \cap A)$, Eq. \eqref{eq:CondProbab} can be recast as follows:
\begin{equation}
\label{eq:CondProbabI}
    P(A \cap B) = P(A|B)P(B) = P(B|A)P(A).
\end{equation}
Bayes’ theorem can simply be obtained by dividing the right-hand side of Eq. \eqref{eq:CondProbabI} by $P(B)$, which gives:
\begin{equation}
\label{eq:BayesTheorem}
    P(A|B) = \frac{P(B|A)P(A)}{P(B)}.
\end{equation}
Bayes’ theorem, as expressed in Eq. \eqref{eq:BayesTheorem}, connects the conditional probability $P(B|A)$ to other conditional probability $P(A|B)$. $P(A)$ is the prior probability and $P(A|B)$ stands for posterior probability, which are two basic concepts specific to Bayesian methods. While the posterior is the conditional probability based on the event B, the prior is unconditioned probability which is used to integrate the expert knowledge or previous experience.

\subsection{Bayesian inference}
\label{BayesInfer}
Bayesian statistics makes use of Bayes theorem, described earlier, to fit a statistical model to the problem at hand. This is achieved by updating the prior knowledge on hyper-parameters $\Theta$, defined by the conditional probability $\Theta \sim P(\Theta)$, with possibly few observation data points. These hyper-parameters are treated as random variables, whose subjective definition must disclose the available information \textit{before} any measurement of the quantity of interest, $\mathcal{Y}$, is performed.

Bayesian theorem in the context of the statistical inference can be recast as the following:
\begin{equation}
\label{eq:BayesInference}
    P(\theta|\mathcal{Y}) = \frac{p(\mathcal{Y}|\theta)P(\theta)}{P(\mathcal{Y})},
\end{equation}
where $P(\theta|\mathcal{Y})$ denotes the posterior distribution of the hyper-parameters, $p(\mathcal{Y}|\theta)$ the likelihood function and $P(\mathcal{Y})$ is the probability of the data.

Assuming that a measured data set of $\mathcal{Y}=\left(\mathcal{Y}_1, ..., \mathcal{Y}_N\right)^T$ with the independent realization is available, the probability of observing the data can be defined by the likelihood function as the following:
\begin{equation}
\label{eq:BayesLikelihood}
    \begin{split}
    p(\mathcal{Y}|\theta) &:= \prod_{i=1}^{N}p(\mathcal{Y}_i|\theta)\\
    &=\frac{1}{\sqrt{(2 \pi)^{N} \operatorname{det} \Sigma}} \exp \left(-\frac{1}{2}(\mathcal{M}(\theta)-\mathcal{Y})^{T} \Sigma^{-1}(\mathcal{M}(\theta)-\mathcal{Y})\right),
    \end{split}
\end{equation}
where $\Sigma$ denotes the covariance matrix, which includes the measurement error as well as all other error sources. 
$P(\mathcal{Y})$, in Eq. \eqref{eq:BayesInference}, is a normalization factor, which ensures that the posterior probability sums up to one. It is also known as \textit{marginal likelihood} or \textit{evidence} and can take the following form: 
\begin{equation}
\label{eq:BayesEvidence}
    P(\mathcal{Y}) = \int_{\Theta} p(\mathcal{Y}|\theta)P(\theta) d\theta.
\end{equation}

The posterior distribution in Eq. \eqref{eq:BayesInference} provides a summary of the inferred information regarding the hyper-parameters after updating the prior knowledge with the observed data. However, the practical computation of the posterior distribution $P(\theta|\mathcal{Y})$ is nothing but trivial. There exists analytical expressions only for special choices of the prior distribution $P(\theta)$. For more involved cases, one approach is to use sampling methods, such as Monte Carlo, or Markov Chain Monte Carlo sampling methods to approximate the posterior distribution.

One advantage of using the Bayesian inference in the task of model validation is that the uncertainty on hyper-paramters $\theta$ can be incorporated into prior and posterior evaluations of quantities of interest $y$ by so-called \textit{predictive distributions}. The prior predictive distribution can be obtained by the following expression:
\begin{equation}
\label{eq:PriorPredictive}
    P'_{pred}(y) := \int_{\Theta} p(y|\theta)P(\theta) d\theta,
\end{equation}
which is average of the conditional distribution of the prior distribution over the prior distribution $P(\theta)$. Similarly, the posterior predictive distribution $P''_{pred}(y|\mathcal{Y})$ can be computed by averaging the conditional distribution $P(y|\theta)$ over the posterior parameter distribution $P(\theta|\mathcal{Y})$, defined in Eq. \eqref{eq:BayesInference}:
\begin{equation}
\label{eq:PosteriorPredictive}
    P''_{pred}(y) := \int_{\Theta} p(y|\theta)P(\theta|\mathcal{Y}) d\theta = \frac{1}{P(\mathcal{Y})} \int_{\Theta} p(y|\theta) p(\mathcal{Y}|\theta) P(\theta) d\theta.
\end{equation}

\section{Model comparison in Bayesian framework}
\label{BayesianModelComparison}
To compare alternative computational models with possibly distinct conceptual models, various strategies have been suggested in the literature. The benefits of this comparison are twofold. First, this evaluates their strengths and weaknesses. Second, their predictive ability is assessed. Hoeting et al. \cite{hoeting1999bayesian} proposed Bayesian model averaging (BMA) as a formal statistical approach, which allows comparing alternative conceptual models, testing their adequacy, combining their predictions into a more robust output estimate, and quantifying the contribution of conceptual uncertainty to the overall prediction uncertainty.

The BMA method is grounded on Bayes’ theorem, which, as mentioned earlier, combines a prior belief about the efficacy of each model with its performance in replicating a common measurement data set. BMA can be regarded as a Bayesian hypothesis testing framework, combining the idea of classical hypothesis testing with the ability to examine multiple alternative models against each other in a probabilistic manner.
It returns model weights that represent posterior probabilities for each model to be the most appropriate one from the set of proposed competing models. Additionally, the computed weights can provide a ranking and a quantitative comparison of the competing models.
Bayes’ theorem closely follows the principle of parsimony or Occam’s razor \cite{angluin1983inductive}, in that the posterior model weights offer a compromise between model complexity and goodness of fit, also known as the bias-variance trade-off \cite{geman1992neural}.

\subsection{Bayesian Model Averaging Framework}
Let us consider that $N_m$ plausible, competing models $M_k$ are available. The posterior predictive distribution of a quantity of interest $\theta$ in Eq. \eqref{eq:PosteriorPredictive} given the vector of observed data $\mathcal{Y}$ can be expressed as:
\begin{equation}
\label{eq:PostPredictiveBMA}
    P''_{pred}(\theta|\mathcal{Y}) := \sum_{k=1}^{N_m} p(\theta|\mathcal{Y}, M_k) P(M_k|\mathcal{Y}),
\end{equation}
where $P(M_k|\mathcal{Y})$ being discrete posterior model weights.The weights can be interpreted as the Bayesian probability of the individual models to be the best representation of the system from the pool of competing models.
The model weights (posterior probabilities of models) are given by Bayes’ theorem, which can be recast for a set of $M_k$ competing models as:
\begin{equation}
\label{eq:BayesBMA}
    P(M_k|\mathcal{Y}) = \frac{p(\mathcal{Y}|M_k) P(M_k)}{\sum_{i=1}^{N_m} p(\mathcal{Y}|M_i) P(M_i)}, 
\end{equation}
where $P(M_k)$ denotes the prior probability, also known as the subjective model credibility that model $M_k$ could be the the most plausible model in the set of models \textit{before} any comparison with observed data have been made.
Hoeting et al. \cite{hoeting1999bayesian} proposed that a "reasonable, neutral choice" could be equally likely priors, i.e. $P(M_k)=1/N_m$, in case of paucity of prior knowledge regarding the merit of the different models under study. The denominator in Eq.\eqref{eq:BayesBMA} is the normalizing constant of the posterior distribution of the models and can simply be obtained by determination of the individual weights. 
Since all model weights are normalized by the same constant, this normalizing factor could even be neglected. Thus, the ranking of the individual models against each other can be represented by the proportionality:
\begin{equation}
\label{eq:Proportionality}
    P(M_k|\mathcal{Y}) \propto p(\mathcal{Y}|M_k) P(M_k).
\end{equation}
%
$p(\mathcal{Y}|M_k)$ expresses the Bayesian model evidence (BME) term, as introduced in Section \ref{BayesInfer}. BME is also referred to as marginal likelihood or prior predictive, since it quantifies the likelihood of the observed data based on the prior distribution of the parameters. 
The BME term can be estimated by integration over the full parameter space $\Theta_k$, which is known as Bayesian integral by Kass and Raftery \cite{kass1995bayes}, and is expressed as:
\begin{equation}
\label{eq:BME}
    p(\mathcal{Y}|M_k) = \int_{\Theta_k} p(\mathcal{Y}|M_k,\theta_k) P(\theta_k|M_k) d\theta_k,
\end{equation}
with $\theta_k$ being the parameter vector of model $M_k$ with the dimension of $N_{p,k}$. $\Theta_k$ denotes the parameter space of model $M_k$, and $P(\theta_k|M_k)$ is the corresponding prior distribution. 
The likelihood or probability of the parameter set $\theta_k$ of model $M_k$ to have generated the measurement data set is represented by $p(\mathcal{Y}|M_k,\theta_k)$ in Eq. \eqref{eq:BME}. For more details on the properties of BME and a comparison of available techniques to evaluate this term, the reader is referred to Schöniger et al. \cite{schoniger2014model}. Here,we perform a brute-force MC integration over each model’s parameter and input space to obtain the BME values.

\subsection{Bayesian hypothesis testing}
\label{BayesHypoTest}
The Bayesian hypothesis testing framework was first introduced by Jeffreys \cite{jeffreys1961theory}. The key component in this framework is the Bayes factor, whose extensive description in the context of practical applications is given by the review paper of Kass and Raftery \cite{kass1995bayes}. They define the Bayes factor as the ratio of BME for two alternative models. Stated differently, Bayes factor, $BF(M_k , M_l)$, can be interpreted as ratio between the posterior and prior odds of model $M_k$ being the more plausible one in comparison to the alternative model $M_l$ :
\begin{equation}
\label{eq:BayesFactor}
    BF(M_k , M_l) = \frac{P(M_k|\mathcal{Y})}{P(M_l|\mathcal{Y})} \frac{P(M_l)}{P(M_k)} = \frac{p(\mathcal{Y}|M_k)}{p(\mathcal{Y}|M_l)}.
\end{equation}

The Bayes factor is regarded as a measure for significance in Bayesian hypothesis testing. It quantifies the evidence (literally, as in Bayesian model evidence) of hypothesis $M_k$ against the null-hypothesis $M_l$.
Jeffreys provided a rule of thumb in his book, Theory of probability \cite{jeffreys1961theory} for the interpretation of Bayes factor values on a log 10-scale. The grades of evidence is summarized in Table \ref{tab:BayesFactorThresholds}.

\begin{table}[h]
\caption{Interpretation of Bayes Factor in favor of model $M_K$ according to \cite{jeffreys1961theory}}
\label{tab:BayesFactorThresholds}
\centering
\begin{tabular}{ll}
\hline\noalign{\smallskip}
$\log_{10}(BF)$ & Interpretation  \\ 
\noalign{\smallskip}\hline\noalign{\smallskip}
1 - 3  & anecdotal evidence \\
3 - 10  & substantial evidence \\
10 - 100  & strong evidence \\
$>$ 100  & decisive evidence \\
\noalign{\smallskip}\hline
\end{tabular}
\end{table}

Following this suggestion, a Bayes factor which lies between 1 and 3 indicates evidence in favor of $M_k$ that is "not worth more than a bare mention", a factor of up to 10 represents "substantial" evidence, and a factor between 10 and 100 can be regarded "strong" evidence. Finally, a Bayes factor greater than 100 admits "decisive" evidence, i.e., it can be used as a threshold to reject models based on poor performance in comparison to the best performing model in the set.

\subsection{Theoretical Upper Limit for Model Performance}
\label{TOM}
Bayes Factor in the context of Bayesian hypothesis testing provides a performance comparison of pairwise competing models. However, in a validation benchmark task, we are also interested in comparing their performance to the best achievable performance.
Schöneger et al. \cite{schoniger2015statistical} argued that this theoretical upper limit for model performance exists when the measurement data set has noise. They propose that this limit can be established via determining a distribution of BME for a so-called \textit{theoretically optimal model} (TOM), which is also dubbed as a sure-thing hypothesis by Jaynes \cite{jaynes2003probability}. They define the observed data set as TOM, as it gives an exact fit with zero bias while having a minimum number of parameters, i.e. exactly zero, which is equivalent to zero variance.
Stated differently, the TOM indicates the expected best possible performance in presence of measurement error. This theoretically optimal distribution of BME in presence of measurement noise can be defined as the distribution of likelihoods of the observed data set given the perturbed data sets. Assuming that measurement errors follow a Gaussian distribution and are independent and identically distributed, the TOM performance (shown as log-BME) has a distribution of the weighted sum of normal squared residuals. Consequently, this distribution can be defined by the chi-square distribution \cite{hald1998history} as:
\begin{equation}
\label{eq:TOMBME}
    \chi^2(x) = \frac{1}{2^(k/2)\Gamma(k/2)} x^{k/2 -1} \exp{(-x/2)},
\end{equation}
with $k$ being the number of degrees of freedom, which is equal to the size of the observed data set $N_s$.

%% file: TEXs/PCE.tex
\chapter{Surrogate Modeling}
\thispagestyle{empty}

\section{Uncertainty Propagation in Bayesian Analysis}
A classic Bayesian analysis, discussed in the previous chapter, requires propagation of the parametric uncertainty through the given computational model. This task is also known as the \textit{uncertainty propagation} (UP). Typically, a significant number of model evaluations are needed in order to yield convergent statistics. In practice, however, the computational complexity of the underlying computational model, as well as the total available computational budget severely restrict the number of evaluations that one can actually carry out. In such situations, the estimates produced by the Bayesian analysis lack sufficient trust, as the limited number of model evaluations can yield additional uncertainty.

The most common approach when dealing with expensive models is to replace them with easy-to-evaluate surrogates. Simply put, one evaluates the model on a set of design points and then strives to establish an accurate relationship between the response surface and the design points. Then, the original computational model can be substituted by its surrogate in the Bayesian analysis. The  polynomial chaos  expansion  (PCE) is one of the most rigorous approach to UP, thanks to its strong mathematical basis and ability to provide functional representations of stochastic quantities.
However, the accuracy of the prediction of these surrogate models, trained with only a handful of simulations is debatable. This argument is rooted in the fact that the surrogates do not attempt to quantify the epistemic uncertainty associated with their predictions.

The goal of this chapter is to highlight how a surrogate model using a PCE can be constructed for computationally intensive models with as few simulations as the computational budget allows. It is also shown how the Bayesian formalism can be materialized by employing the concept of PCE to account for the uncertainty in surrogate's predictions. Moreover, we introduce a set of sequential adaptive sampling strategies, in which one attempts to augment the initial design in an iterative manner. Doing so, interesting regions in the parameter space are properly explored, avoiding the waste of computational resources as opposed to the so-called one-shot designs. These regions are more likely to provide valuable information on the behavior of the original model responses.

\section{Polynomial Chaos Expansion}
In a probabilistic framework, uncertainties in input parameters are modeled via random variables. These input uncertainties can be investigated using a Polynomial chaos expansion (PCE). This method provides the means to develop an approximation to the map between inputs and the quantities of interest (QoI). This mapping is both computationally tractable and sufficiently accurate. The main idea of PCE is to expand a  QoI with a finite variance in a suitably built basis of multivariate polynomials that are orthogonal to the joint probability density functions of the inputs. It is worth noting that the random variables are assumed to be statistically independent or may be linearly correlated. The linear correlation can be handled by adequate linear transformation \cite{oladyshkin2012data}.

A PCE is a linear regression that includes linear combinations of a fixed set of nonlinear functions with respect to the input variables, known as basis functions (Section \ref{BasisFunc}). The PCE of the random variable $\mathrm{Y}$ can be cast as the following:
\begin{equation}
\label{Eq:PCE_Gen}
    \mathrm{Y} = \mathcal{M}(\mathbf{X}) = \sum_{\bm{\alpha} \in \mathbb{N}^M} c_{\bm{\alpha}} \Psi_{\bm{\alpha}}(\mathbf{X}),
\end{equation}
where $\Psi_{\bm{\alpha}}(\mathbf{X})$ represents multivariate polynomials orthogonal with respect to $f_{\mathbf{X}}$ and $\bm{\alpha}$ denotes a multi-index that represents the components of the multivariate polynomials $\Psi_{\bm{\alpha}}$. The $c_{\bm{\alpha}} \in \mathbb{R}$ are the corresponding coefficients (coordinates).
For practical reasons, the sum in Eq. \eqref{Eq:PCE_Gen} needs to be truncated to a finite sum, by introducing the truncated polynomial chaos expansion:
\begin{equation}
\label{Eq:PCE_Trunc}
    \mathcal{M}(\mathbf{X}) \approx \mathcal{M}^{PC}(\mathbf{X}) = \sum_{\bm{\alpha} \in \mathcal{A} } c_{\bm{\alpha}} \Psi_{\bm{\alpha}}(\mathbf{X}),
\end{equation}
where $\mathcal{A} \subset \mathbb{N}^M$ denotes the set of selected multi-indices of the multivariate polynomials. A standard truncation scheme can be defined as all polynomials in the $M$ input variables of total degree less or equal to $p$:
\begin{equation}
\begin{split}
\label{Eq:truncation}
    \mathcal{A}^{M, p} = \{ \bm{\alpha} \in \mathbb{N}^M \ : \ |\bm{\alpha}|\leq p\} \\
    \text{card} \ \mathcal{A}^{M, p} \equiv P =\binom{M+p}{p}
\end{split}
\end{equation}
For more other truncation schemes, the reader is referred to \cite{marelli2015uqlab}.
\subsection{Polynomial basis functions} 
\label{BasisFunc}
The multivariate polynomials $\Psi_{\bm{\alpha}}(\mathbf{X})$ are tensor product of the univariate polynomials:
\begin{equation}
\label{Eq:Psi}
    \Psi_{\bm{\alpha}}(\mathbf{X}) :=  \prod_{i=1}^M \psi_{\alpha_i}^{(i)}(x_i).
\end{equation}
The univariate orthonormal polynomials $\psi_{\alpha_i}^{(i)}(x_i)$ must satisfy the following:
\begin{equation}
\label{Eq:univPsi}
    \langle \psi_j^{(i)}(x_i), \psi_k^{(i)}(x_i) \rangle := \int_{\mathcal{D}_{X_i}} \psi_j^{(i)}(x_i) \psi_k^{(i)}(x_i) f_{X_i} \ (x_i)d x_i = \delta_{j k}.
\end{equation}
where $i$ represents the input variable with respect to which they are orthogonal as well as the corresponding polynomial family, $j$ and $k$ the corresponding polynomial degree, $f_{X_i}(x_i)$ is the $i$th-input marginal distribution and $\delta_{j k}$ is the Kronecker delta.
The classical families of univariate orthonormal polynomials are given for reference in \cite{sudret2007uncertainty}. For detailed description of each of this classical families, the reader is referred to \cite{xiu2002wiener}.

The calculation of polynomial basis via the classical families is grounded in the fact that an exact knowledge about the probability density functions is available. However, the information about the distribution of the data is distinctly restricted in engineering applications, most importantly when environmental influences or natural phenomena are of interest or when prediction is involved. For instance, the material properties of subsurface reservoirs are not readily available to shed light on their distribution.
Oladyshkin \& Nowak \cite{oladyshkin2012data} demonstrate that statistical moments are the only source of information that is propagated in all polynomial expansion-based stochastic approaches. The author leverage this fact to propose an arbitrary polynomial chaos expansion (aPCE), that can operate with probability measures that may be implicitly and incompletely defined via their statistical moments. Using aPCE, one can build the orthonormal polynomials even in the absence of the exact probability density function $f_{\mathbf{X}}(x)$.
\subsection{Calculation of the coefficients using Bayesian sparse learning}
Recently, \textit{sparse} learning algorithms in the context of linear modeling has received attention. These algorithms set many weights to zero in the estimator predictor function. Sparsity is an attractive concept, which offers elegant complexity control, over-fitting control, feature extraction, the potential for characterization of meaningful input variables along with the practical computational speed and compactness. As stated earlier, PCEs also belong to linear regression models and employing the concept of sparsity can lead to zero values for many $c_{\bm{\alpha}}$ to in the expansion in Eq. \eqref{Eq:PCE_Trunc}.

There are many mathematical approaches when dealing with a regression problem that lead to a sparse solution. These approaches have led to the emerge of numerous sparse solvers in the compressed sensing (e.g. \cite{arjoune2017compressive}), as well as in the sparse PCE. The proposed solvers in the context of PCE can be categorized into convex optimization solvers, greedy methods, iteratively re-weighted methods, and Bayesian sparse learning, which is also known as compressive sensing. For further details on different solvers in each category, the reader is referred to the comprehensive survey of L\"uthen et al. \cite{lthen2020sparse}. Here, we employ sparsity within a Bayesian framework by a Bayesian sparse learning method to be able to provide a probabilistic prediction, i.e. a prediction with the associated uncertainty. This prediction uncertainty can be used as the expected error occurring when replacing the original computational model by a possibly less accurate surrogate.

In Bayesian sparse learning, one imposes a sparsity-inducing prior on the coefficients (weights) of the predictors ($\Psi_{\alpha_i}$ in the expansion \eqref{Eq:PCE_Trunc}), whose parameters are considered to be random variables with a hyperprior. Then, the posterior of the weights are inferred e.g. using a fast marginal likelihood maximization algorithm \cite{tipping2003fast}. This learning process leads to extremely sparse inferred predictors, since they yield relatively few non-zero $c_i$ parameters. That means a significant number of the predictors give posterior distributions centered at zero.

Let the target variable be $\mathrm{Y}$, which is given by a deterministic function $y(\mathbf{X},\mathbf{c})$ with an additive Gaussian noise which reads:
\begin{equation}
\label{Eq:target}
    \mathrm{Y} = y(\mathbf{X},\mathbf{c}) + \epsilon,
\end{equation}
where $\epsilon$ is a zero mean Gaussian random variable with precision (inverse variance) $\beta$.
Hence, the equation can be cast as:
\begin{equation}
\label{Eq:ML_Linear_Prob}
    p(\mathrm{Y}|\mathbf{X},\mathbf{c}, \beta) = \mathcal{N}( \mathrm{Y}|y(\mathbf{X},\mathbf{c}) , \beta^{-1} ).
\end{equation}
Let $\mathbf{X}=\{ x_1, ..., x_N \}$ be a data set of inputs with the corresponding model responses $\mathrm{Y} = \{ \mathcal{Y}_1, ..., \mathcal{Y}_N \}$. One can group the model responses into a column vector denoted by $\mathrm{\textbf{Y}}$ to be distinguished from a single observation of a multivariate response, which would be denoted by $\mathrm{Y}$. Assuming these data points are drawn independently from the distribution in Eq.\eqref{Eq:ML_Linear_Prob}, and using Eq.\eqref{Eq:PCE_Gen}, a multivariate Gaussian likelihood function can be derived as:
\begin{equation}
\label{Eq:PCE_likelihood}
\begin{split}
    p(\mathrm{\mathbf{Y}}|\mathbf{X},\mathbf{c}, \beta) &= \prod_{n=1}^{N} \mathcal{N}( y_n|\mathbf{c}^\top \Psi (\mathbf{x}_n) , \beta^{-1} ) \\
    & =  \left({2 \pi \beta^{-1}}\right)^{-N/2} \exp \left\{-\frac{\beta}{2} {||y_n - \mathbf{c}^\top \Psi (\mathbf{x}_n)||}^2 \right\},
\end{split}
\end{equation}
where this is a function of the parameter $\mathbf{c}$ and $\beta$.

We introduce a Gaussian prior distribution over the parameter vector $\mathbf{c}$ by giving each of the weight parameters $c_i$ a separate hyper-parameter $\alpha_i$. Thus, the prior of the weights reads as:
\begin{equation}
\label{Eq:PCE_Prior}
\begin{split}
    p(\mathbf{c}|\bm{\alpha}) &= \prod_{i=1}^{P} \mathcal{N}( c_i|0, \alpha_i^{-1} )\\
    &=\prod_{i=1}^{P} \left[{2 \pi}^{-1/2} {\alpha_i}^{1/2} \exp \left\{-\frac{1}{2} \alpha_i c^2_i \right\} \right],
\end{split}
\end{equation}
where $\bm{\alpha}=\{ \alpha_i, ..., \alpha_M\}^\top$ denotes the precision of the prior over its associated weight parameter $\mathbf{c}$. The form of prior is ultimately responsible for the sparsity properties of the model (for more details, see \cite{tipping2001sparse}).
The posterior distribution, conditioned on the model responses, is given by combining the likelihood in Eq. \eqref{Eq:PCE_likelihood} and the prior Eq. in \eqref{Eq:PCE_Prior} according to Bayes' rule. This posterior, given $\bm{\alpha}$, can take the form:
\begin{equation}
\label{Eq:PCE_Posterior}
    p(\mathrm{\mathbf{c}}|\mathbf{Y},\bm{\alpha}, \beta) = \frac{p(\mathrm{\mathbf{Y}}|\mathbf{X},\mathbf{c}, \beta) p(\mathbf{c}|\bm{\alpha})}{p(\mathrm{\mathbf{Y}}| \mathbf{X}, \bm{\alpha} , \beta)},
\end{equation}
which is Gaussian $\mathcal{N}( \mathbf{c}| \bm{\mu}, \mathbf{\Sigma})$ with:
\begin{equation}
\label{Eq:PCE_Posterior_moments}
\begin{split}
    \bm{\mu} &= \beta \mathbf{\Sigma} \mathbf{\Psi}^\mathcal{T} \mathrm{\mathbf{Y}} \\
    \mathbf{\Sigma} &= \left(\mathbf{A}+ \mathbf{\Psi}^\mathcal{T} \beta \mathbf{\Psi} \right)^{-1} ,
\end{split}
\end{equation}
where $\mathbf{\Psi}$ is the design matrix of the size $N \times M$ with elements $\Psi_{ni}=\psi_i(x_n)$, and $\mathbf{A}=\mathrm{diag}(\alpha_i)$. The values of $\bm{\alpha}$ and $\beta$ can be determined via type-II maximum likelihood \cite{berger2013statistical}, also known \textit{evidence approximation} in the machine learning literature \cite{gull1989developments,mackay1992bayesian}.

\subsection{Prediction with Bayesian sparse PCE}
\label{PredictBaSPCE}
Having found values $\bm{\alpha}^*$ and $\mathbf{\beta}^*$ for the hyperparameters that maximize the marginal likelihood, one can evaluate the predictive distribution over $\mathrm{Y}$ for a new input $\mathrm{\mathbf{x}}$ by:
\begin{equation}
\label{Eq:Pred_Dist}
\begin{split}
    p(\mathrm{Y}| \mathbf{x}, \mathbf{X}, \mathrm{\mathbf{Y}}, \bm{\alpha}^* , \beta^*) &= \int p(\mathrm{Y}| \mathbf{x}, \mathbf{c} , \beta^*) p(\mathbf{c}|\mathbf{X}, \mathrm{\mathbf{Y}}, \bm{\alpha}^*, \beta^*) d\mathbf{c} \\
    &= \mathcal{N}(\mathrm{Y} | \mathbf{\mu}^\top \Psi(\mathbf{x}), \sigma^2(\mathbf{x})).
\end{split}
\end{equation}
The predictive mean is given by Eq.\eqref{Eq:PCE_Posterior_moments} with $\mathbf{c}$ set to the posterior mean $\bm{\mu}$, and the variance of the predictive distribution is given by:
\begin{equation}
\label{Eq:Pred_Dist_sigma2}
    \sigma^2(\mathbf{x}) = (\beta^*)^{-1} + \Psi(\mathbf{x})^\top \mathbf{\Sigma} \Psi(\mathbf{x})
\end{equation}
where $\mathbf{\Sigma}$ is calculated by Eq.\eqref{Eq:PCE_Posterior_moments} in which $\bm{\alpha}$ and $\beta$ set to their optimized values $\bm{\alpha}^*$ and $\beta^*$.


\section{On the accuracy of the surrogate model}
Once the surrogate model is constructed, its accuracy and predictive capability need to be assessed. In case, a $N_v$ independent set of inputs and outputs, a.k.a a validation set $\left[ (x^{(1)}, \mathcal{M}(x^{(1)})), ..., (x^{(j)}, \mathcal{M}(x^{(j)}))\right]$, is available in addition to a training set for training the surrogate model, the \textit{validation error} can be computed as:
\begin{equation}
\label{Eq:ValidationError}
    \epsilon_{Val} = \frac{N_v-1}{N_v} \left[ \frac{\sum^{N_v}_{i=1} \big({\mathcal{M}(x^{(i)}_{Val})-\mathcal{M}^{PC}(x^{(i)}_{Val})\big)}^2}{\sum^{N_v}_{i=1}\big({\mathcal{M}(x^{(i)}_{Val})-\hat{\mu}_{Y_{Val}}\big)}^2} \right],
\end{equation}
where $\hat{\mu}_{Y_{Val}}=\frac{1}{N_v}\sum^{N_v}_{i=1}\mathcal{M}(x^{(i)}_{Val})$ denotes the mean of model responses for the validation set. Since computation of the aforementioned error requires a large number of model evaluations, it is only computationally tractable for simple analytical functions. To avoid additional model evaluations for assessing the accuracy of the surrogate model, an error based on the already evaluated ED is more desirable.

One common approach is the leave-one-out error (LOO), proposed by \cite{geisser1975predictive,stone1974cross} explicitly introduced for PCE. This error, denoted by $\epsilon_{LOO}$, is composed of rebuilding $N$ surrogate models in sequential ($\mathcal{M}^{PC \setminus i}$), using the original experimental design excluding $i$-th set ($\mathbf{X} \setminus \mathbf{x}^{(i)}$). Then the prediction error at the excluded set ($\mathbf{x}^{(i)}$) is computed. For more details, see \cite{blatman2011adaptive}.
Blatman \cite{blatman2009adaptive} shows that calculating $N$ independent surrogates is not needed, when using the linear superimposition of orthogonal terms, which is the case for PCE. Alternatively, the error can be calculated analytically from a single surrogate based on all sets in the ED using the following equation:
\begin{equation}
\label{Eq:LOOCV2}
    \epsilon_{LOO} = \frac{\sum_{i=1}^{N} {\left( \frac{\mathcal{M}(x^{(i)}) - \mathcal{M}^{PC}(x^{(i)})}{1-h_i}  \right)}^2}{\sum_{i=1}^{N} {\left( \mathcal{M}(x^{(i)}) - \mu_\mathcal{M}  \right)}^2} ,
\end{equation}
where $h_i$ is the $i$-th diagonal entry of the experimental matrix $\mathcal{A}(\mathcal{A}^{\mathcal{T}}\mathcal{A})^{-1} A^\mathcal{T}$.

\section{Experimental design}
The computational cost of construction of a surrogate model and its accuracy crucially depends on the number of required evaluations of the computationally expensive forward model on the so-called experimental design (ED). ED is a set of training samples from the joint distribution of the input parameters. Properly designed ED has proved vital for simultaneous reduction of the effect of noise and bias errors which can raise the confidence in the task of Bayesian analysis. 
This has motivated researchers to examine assorted strategies for constructing the sample set $\{\Psi^{(i)} \}^N_{i=1}$ beyond the standard MC sampling.
In this context, the influence of different experimental designs on predictions have been adequately addressed in the literature  \cite{simpson2001sampling,giunta2003overview,simpson2004approximation,queipo2005surrogate,fajraoui2017sequential,hadigol2018least}.

In general, the sampling approaches can be categorized into two groups: \textit{classical sampling} and \textit{sequential sampling}. 
The common practice in classical (one-shot) sampling is to choose the experimental design $P$ grounded only in the information that is available prior to any model evaluation, e.g. existence of noise, relevance of the input parameters, measurement precision. Then, the computational model is evaluated on the selected samples in the ED, and the surrogate model is finally created. This approach is also known as one-shot approach, as all the sample points in the ED are specified at once and no later evaluations of additional samples are made. This is a quite challenging task, since the determination of an optimal sample size is hindered by lack of prior knowledge about the model behavior.

To tackle this problem, flexible sequential sampling strategies have been proposed, which sequentially determine the samples in the design using the information from previous iterations.
The sequential sampling approaches can be grouped into two categories: \textit{space-filling sequential sampling} and \textit{sequential adaptive sampling}. Space-filling approaches make sure that the generated samples cover over the entire domain evenly. These sampling approaches are usually developed from some one-shot sampling criteria by selecting the points in a sequential manner \cite{fajraoui2017sequential}. 
However, the adaptive sequential sampling, also known as adaptive sampling and active learning in machine learning \cite{settles2009active}, makes more informed choices of samples via the surrogate model itself or data that it learns from, and hence, achieves better performance with fewer points than the space-filling sampling, resulting in saving the simulation cost of expensive computational models \cite{liu2018survey}.

Here, we adopt a sequential adaptive sampling experimental design (SAED).
Firstly, an initial batch of samples is selected via a one-shot experimental design. This design can be produced by a common DOE approach, such as Latin Hyper-cube sampling or random sampling. Next, the model is evaluated provided the previously selected samples. Then, the surrogate model is trained to construct a relationship between the ED and the quantities of interest.      
After construction of the surrogate model, its accuracy is estimated using an error metric, e.g. Leave-one-out error in Eq. \eqref{Eq:LOOCV2}. Given that the initial ED is chosen to be small, the estimated error metric most probably suggests that the enrichment of ED is needed until either the accuracy requirement is met or the maximum allowed total number of runs is reached. Algorithm \ref{alg: SeqDesign} summarizes a typical sequential experimental design method.
\begin{algorithm}[ht]
\caption{A typical sequential design method}
\label{alg: SeqDesign}
\SetAlgoLined
\KwResult{Enriched experimental design}
 $P \leftarrow$ initial experimental design\;
 Evaluate the computational model at $P$\;
 Train the surrogate model\;
 Compute the error metric\;
 \While{error $>$ prescribed error or No. runs $<$ Total No. runs}{
  Select new sample $P_{new}$ using sequential design strategy\;
  Evaluate the computational model at $P_{new}$\;
  $P \leftarrow P \cup P_{new}$\;
  Train the surrogate model\;
 }
\end{algorithm}
%

The selection of these additional samples is made by an active learning sampling strategy. Some of these strategies will be explained later in detail. Finally, a new surrogate model is built using all the data gathered thus far, and the model accuracy is estimated again. If either the prescribed accuracy level of the surrogate model or total number of samples is still not reached, the entire sample selection process is repeated. Through sequential selection of samples, more information is available to improve sampling, compared to classical design of experiments \cite{crombecq2011surrogate}. 
The ultimate goal of this algorithm is to reduce the overall number of samples, as evaluating the samples (running the simulation) is the dominant cost in the entire surrogate modelling process. Essentially, non-optimal designs require more resources to make inferences on the features of interest with the same (or less) level of reward that an optimal design would. Hence, when the computational bottleneck is the evaluation of the QoI for any given realization, the additional computational cost of constructing an optimal design is justifiable.

In what follows, the learning strategies (design criteria) for sequential enrichment of the ED are introduced. These criteria can be implemented for emulators being trained for model calibration, i.e. the measurement data is available and can assist in the task of Bayesian model validation.
In Section \ref{BayesActLearning}, a Bayesian optimal design is introduced for sequential learning method. The objective is to identify a new design, among some prospective designs, which maximizes the expected utility.
Moreover, these strategies are investigated with an analytical example with 10 input parameters. 

\subsection{Bayesian active learning}
\label{BayesActLearning}
The Bayesian framework provides a principled approach for incorporating prior information and/or uncertainties concerning the statistical model via a utility function which encapsulates the experimental goals. 
Given the current model with design $d$, the optimal next design point $\mathbf{d}^*$ is the one which maximizes the expected utility function $U(\mathbf{d})$ over the design space $\mathcal{D}$ with respect to the future data $y$ and model parameters $\theta$:
\begin{equation}
\label{Eq:BODE}
    \mathbf{d}^* = \argmax\limits_{\mathbf{d} \in \mathcal{D}} \mathrm{U(\mathbf{d})}.
\end{equation}
One needs to consider every possible observation that could be obtained from an experiment with each design, and then evaluate the relative likelihoods and statistical values of these observations. Ultimately, the design that maximizes the expected utility is selected as the optimal design at each stage of experimentation.
In the following sections, we will first show how three choices of utility functions will be introduced. These utilities lead to valid measures of \textit{information gain}, \textit{model evidence} and \textit{information entropy}.

\subsubsection{Bayesian Inference for Bayesian sparse PCE}
\label{Bayesain_BaSPCE}
Bayesian sparse PCE provides predictions as a mean value $\mu_\mathbf{y}(\theta,x,y,z,t)$ and standard deviation $\sigma_\mathbf{y}(\theta,x,y,z,t)$, as discussed in Section \ref{PredictBaSPCE}. Therefore, initial knowledge on model response $\mathbf{y}(\theta,x,y,z,t)$ in each point of space $(x,y,z)$ and time $t$ for the given exploration parameter set $d^+$ from the parameter space $\mathcal{D}$ is encoded in the Gaussian prior probability distribution $\mathcal{N}(\mu_\mathbf{y}(d^+,x,y,z,t),\sigma_\mathbf{y}(d^+,x,y,z,t))$.
Thus, the prior probability distribution of model response $\mathbf{y}(\theta,x,y,z,t)$ for the given parameter set $d^+$ is forming response space $Y$ that is a multivariate Gaussian, denoted as $\mathcal{N}_{d^+}(\mu_\mathbf{y},\sigma_\mathbf{y})$.  According to the Bayesian framework (Section \ref{BayesInfer}), we can obtain a posterior probability distribution $p_{d^+}(\mathbf{y}\vert\mathcal{Y})$ of the model response for the given parameter set $d^+$, incorporating the observed data $\mathcal{Y}$:
\begin{equation}
\label{Eq:BayesTheorem_AL}
	p_{d^+}(\mathbf{y}\vert\mathcal{Y}) = \frac{p_{d^+}(\mathcal{Y}\vert\mathbf{y}) \mathcal{N}_{d^+}(\mu_\mathbf{y},\sigma_\mathbf{y})}{p_{d^+}(\mathcal{Y})}, 
\end{equation}
where the term $p_{d^+}(\mathcal{Y}\vert\mathbf{y})$ is the likelihood function that quantifies how well the surrogate model predictions $\mathbf{y}(d^+,x,y,z,t)$ drawn from the multivariate Gaussian $\mathcal{N}_{d^+}(\mu_\mathbf{y},\sigma_\mathbf{y})$ match the observed data $\mathcal{Y}$ and the term $p_{d^+}(\mathcal{Y})$ denotes the Bayesian model evidence value for the given parameter set $d^+$.

Assuming independent and Gaussian distributed measurement errors, the likelihood function $p_{d^+}(\mathcal{Y}\vert\mathbf{y})$ can be written as:
\begin{equation}
\label{Eq:AL_Likelihood}
p_{d^+}(\mathcal{Y}\vert\mathbf{y}) = (2\pi)^{-\mathrm{N_*}/2}{\vert \mathbf{R}\vert }^{-\frac{1}{2}} \exp \left[-\frac{1}{2} {\left(\mathcal{Y}-\mathbf{y}(d^+,x,y,z,t)\right)}^T \mathbf{R}^{-1} \left(\mathcal{Y}-\mathbf{y}(d^+,x,y,z,t)\right)\right],
\end{equation}
where $\mathbf{y}(d^+,x,y,z,t)$ comes from $ \mathcal{N}_{\bm{\omega}_{e}}(\mu_\mathbf{y},\sigma_\mathbf{y})$ and $N_*$ denotes the number of measurement points.

\subsubsection{Model evidence based utility}
As discussed in Section \ref{BayesianModelComparison}, Bayesian model evidence (BME) can be regarded as a metric to rank competing models. Here, we leverage this property of BME to identify the next suitable training point in the parameter space during the sequential design. In each iteration, we compute the BME value for prospective design point $d^+$ using the following expression:
\begin{equation}
\label{Eq:BME_AL}
\mathrm{BME_{BODE}} \equiv p_{d^+}(\mathcal{Y}) = \int_{Y} p_{d^+}(\mathcal{Y}\vert \mathbf{y}) \mathcal{N}_{d^+}(\mu_\mathbf{y},\sigma_\mathbf{y}) d\mathbf{y}.
\end{equation}
$\mathrm{BME}_{BODE}$ in Equation \eqref{Eq:BME_AL} can be approximated by:
\begin{equation}
\label{Eq:BME_AL_E}
\mathrm{BME_{BODE}}= \mathop{\mathbb{E}_{\mathcal{N}_{d^+}(\mu_\mathbf{y},\sigma_\mathbf{y})}} \left[ p_{d^+}(\mathcal{Y}\vert \mathbf{y}) \right],
\end{equation}
where the term on the right hand side denotes the expected value $\mathop{\mathbb{E}_{\mathcal{N}_{d^+}(\mu_\mathbf{y},\sigma_\mathbf{y})}}$ of the likelihood $p_{d^+}(\mathcal{Y}\vert \mathbf{y})$ over the prior $\mathcal{N}_{d^+}(\mu_\mathbf{y},\sigma_\mathbf{y})$ provided by the surrogate's prediction. Consequently, the next training point for the surrogate model, i.e. $\mathbf{d}^*_{BODE} \in \mathcal{D}$, can be identified by maximizing the model evidence $\mathrm{BME}_{BODE}$, one can find the next training point :
\begin{equation}
\label{Eq:BME_BODE_Optimization}
	 \mathbf{d}^*_{BODE} = \argmax\limits_{d^+ \in \mathcal{D}} \mathrm{{BME}_{BODE}}.
\end{equation}

\subsubsection{Information gain utility}
A utility function based on mutual information is known as one of the most widely used Bayesian design criteria, which is based on relative entropy. This utility function includes Kullback-Leibler divergence \cite{kullback1951information} and seeks to maximize the expected information gain in moving from the multivariate Gaussian prior $\mathcal{N}_{d^+}(\mu_\mathbf{y},\sigma_\mathbf{y})$ to the posterior $p_{d^+}(\mathbf{y}\vert\mathcal{Y})$ during the learning procedure.

Formally,  the relative entropy $\mathrm{DKL_{BODE}} \left[p_{d^+}(\mathbf{y}\vert\mathcal{Y}), \mathcal{N}_{d^+}(\mu_\mathbf{y},\sigma_\mathbf{y}) \right]$ can be defined for each candidate sampling point $d^+$ from the parameter space $\mathcal{D}$ as following:
\begin{equation}
\label{Eq:DKL_BODE}
	 \mathrm{DKL_{BODE}} \left[p_{d^+}(\mathbf{y}\vert\mathcal{Y}), \mathcal{N}_{d^+}(\mu_\mathbf{y},\sigma_\mathbf{y}) \right] = \int_{Y} \ln \left[\frac {p_{d^+}(\mathbf{y}\vert\mathcal{Y})}{\mathcal{N}_{d^+}(\mu_\mathbf{y},\sigma_\mathbf{y})} \right] p_{d^+}(\mathbf{y}\vert\mathcal{Y}) d\mathbf{y}. 
\end{equation}
Following \cite{oladyshkin2019connection}, one can avoid multidimensional integration in Eq. \eqref{Eq:DKL_BODE} by:
\begin{equation}
\label{Eq:DKL_BODE_E}
	 \mathrm{DKL_{BODE}} \left[p_{d^+}(\mathbf{y}\vert\mathcal{Y}), \mathcal{N}_{d^+}(\mu_\mathbf{y},\sigma_\mathbf{y}) \right] =-\ln \mathrm{BME_{BODE}} + \mathop{\mathbb{E}_{p_{d^+}(\mathbf{y}\vert\mathcal{Y})}} \left(  \ln \left[ p_{d^+}(\mathbf{Y}\vert \mathbf{y}) \right] \right).
\end{equation}
Therefore, the optimization problem to select the next training point take the following form:
\begin{equation}
\label{Eq:DKL_BODE_Optimization}
	 \mathbf{d}^*_{BODE} = \argmax\limits_{d^+ \in \mathcal{D}} \mathrm{DKL_{BODE}} \left[p_{d^+}(\mathbf{y}\vert\mathcal{Y}), \mathcal{N}_{d^+}(\mu_\mathbf{y},\sigma_\mathbf{y}) \right].
\end{equation}

It is worth mentioning that Eq. \eqref{Eq:DKL_BODE_Optimization} depends not only on $\mathrm{BME_{BODE}}$ values from Eq. \eqref{Eq:DKL_BODE_E}, but also on the cross entropy represented by term $\mathop{\mathbb{E}_{p_{d^+}(\mathbf{y}\vert\mathcal{Y})}} \left(  \ln \left[ p_{d^+}(\mathcal{Y}\vert \mathbf{y}) \right] \right)$. This term reflects how informative the likelihood is (see details in \cite{oladyshkin2019connection}). Moreover, we obtain the last term via a rejection sampling technique using the evaluations from the already trained surrogate model.

\subsubsection{Information entropy-based utility}
Another utility for selection of the next training point in a BODE has its root in information entropy \cite{shannon1948mathematical} and is often used in machine learning. Here, we aim at reducing the expected information loss during the sequential design. The information entropy $\mathrm{H}_{BODE} \left[ p_{d^+}(\mathbf{y}\vert\mathcal{Y}) \right]$ to asses information loss for each parameter set $d^+$ as the candidate for next training point can be computed by:
\begin{equation}
\label{Eq:Entropy_BODE}
	 \mathrm{H_{BODE}} \left[ p_{d^+}(\mathbf{y}\vert\mathcal{Y}) \right] = - \int_{Y} \ln \left[  p_{d^+}(\mathbf{y}\vert\mathcal{Y}) \right]  p_{d^+}(\mathbf{y}\vert\mathcal{Y}) d\mathbf{y}. 
\end{equation}

According to \cite{oladyshkin2019connection}, information entropy in Eq. \eqref{Eq:Entropy_BODE} can be written as following:
\begin{equation}
\label{Eq:Entropy_BODE_E}
\begin{split}
	  \mathrm{H_{BODE}} \left[ p_{d^+}(\mathbf{y}\vert\mathcal{Y}) \right]  = \ln \mathrm{BME_{BODE}} &- \mathop{\mathbb{E}_{p_{d^+}(\mathbf{y}\vert\mathcal{Y})}} \left( \ln \left[ \mathcal{N}_{d^+}(\mu_\mathbf{y},\sigma_\mathbf{y}) \right] \right) \\ &- \mathop{\mathbb{E}_{p_{d^+}(\mathbf{y}\vert\mathcal{Y})}} \left(  \ln \left[ p_{d^+}(\mathcal{Y}\vert \mathbf{y}) \right] \right).
\end{split}
\end{equation}
We obtain all terms in Eq. \eqref{Eq:Entropy_BODE_E} using prior-bases or posterior-bases sampling on surrogate model's prediction, avoiding any multidimensional integration using methods such as rejecting sampling. Therefore, our optimization problem takes the following form:
\begin{equation}
\label{Eq:Entropy_BODE_Optimization}
	 \mathbf{d}^*_{BODE} = \argmin\limits_{d^+ \in \mathcal{D}} \mathrm{H_{BODE}} \left[ p_{d^+}(\mathbf{y}\vert\mathcal{Y}) \right],
\end{equation}
in that, we seek to identify the parameter set $\mathbf{d}^*_{BODE}$ from the parameter space $\mathcal{D}$ that corresponds to minimum of information entropy $\mathrm{H_{BODE}} \left[ p_{d^+}(\mathbf{y}\vert\mathcal{Y}) \right]$.

\subsection{Numerical experiment}
\label{NumericalExp}
In this study, we consider a non-linear analytical function $\bm{y}(\bm{\theta}, t)$ with ten ($n=10$) uncertain parameter $\bm{\theta}=\{\theta_1, ..., \theta_n \}$, used in \cite{oladyshkin2019connection} as:
\begin{equation}
\label{Eq:AnalyticEq}
    \bm{y}(\bm{\theta}, t) = {(\theta_1^2 + \theta_2 - 1)}^2 + \theta_1^2 + 0.1 \theta_1 \exp{(\theta_2)} - 2 \theta_1 \sqrt{0.5t} + 1 + \sum_{i=2}^n \frac{\theta_i^3}{i},
\end{equation}
where the prior parameter distribution $p(\bm{\theta})$ is considered to be independent and uniform with $\theta_i \sim \mathcal{U}(-5,5)$ for $i=1,...,n$. Moreover, we construct a test scenario by generating ten synthetic observed data values $\bm{y_*} = \bm{y}(\bm{\theta}, t_k)$ with $k=1,...,10$ corresponding to $\theta_i = 0 \ \forall i$. 

To assess the prediction accuracy of $\bm{y}(\bm{\theta}, t)$, in Eq. \eqref{Eq:AnalyticEq} comparing to the synthetic observed data $\bm{y_*}$ , we use the likelihood function in \eqref{eq:BayesLikelihood}, assuming independent and Gaussian distributed error of $\sigma_{\epsilon} = 2$. Due to the random nature of the discussed sampling techniques, the results presented in the following are obtained by running 60 independent replications. As convergence criteria, we monitor the difference of Bayesian model evidence and Kullback-Leibler divergence defined as below with their reference values:
\begin{equation}
\begin{gathered}
\mathrm{BME}=p(\mathbf{D})=\int_{\Theta} p(\mathbf{D} \mid \theta) p(\boldsymbol{\theta}) d \boldsymbol{\theta}=\mathbb{E}_{p(\boldsymbol{\theta})}(p(\mathbf{D} \mid \boldsymbol{\theta})) \approx \frac{1}{N} \sum_{i=1}^{N}\left(p\left(\mathbf{D} \mid \boldsymbol{\theta}_{i}\right)\right) \\
\mathrm{D}_{\mathrm{KL}}[p(\boldsymbol{\theta} \mid \mathbf{D}), p(\boldsymbol{\theta})]=-\ln \mathrm{BME}+\frac{1}{N_{p}} \sum_{i=1}^{N_{p}}\left(\ln \left[p\left(\mathbf{D} \mid \boldsymbol{\theta}_{i}\right)\right]\right)
\end{gathered}
\end{equation}

\begin{figure}[h!]
\begin{subfigure}{1.0\textwidth}
  \centering
  \includegraphics[width=1.0\linewidth]{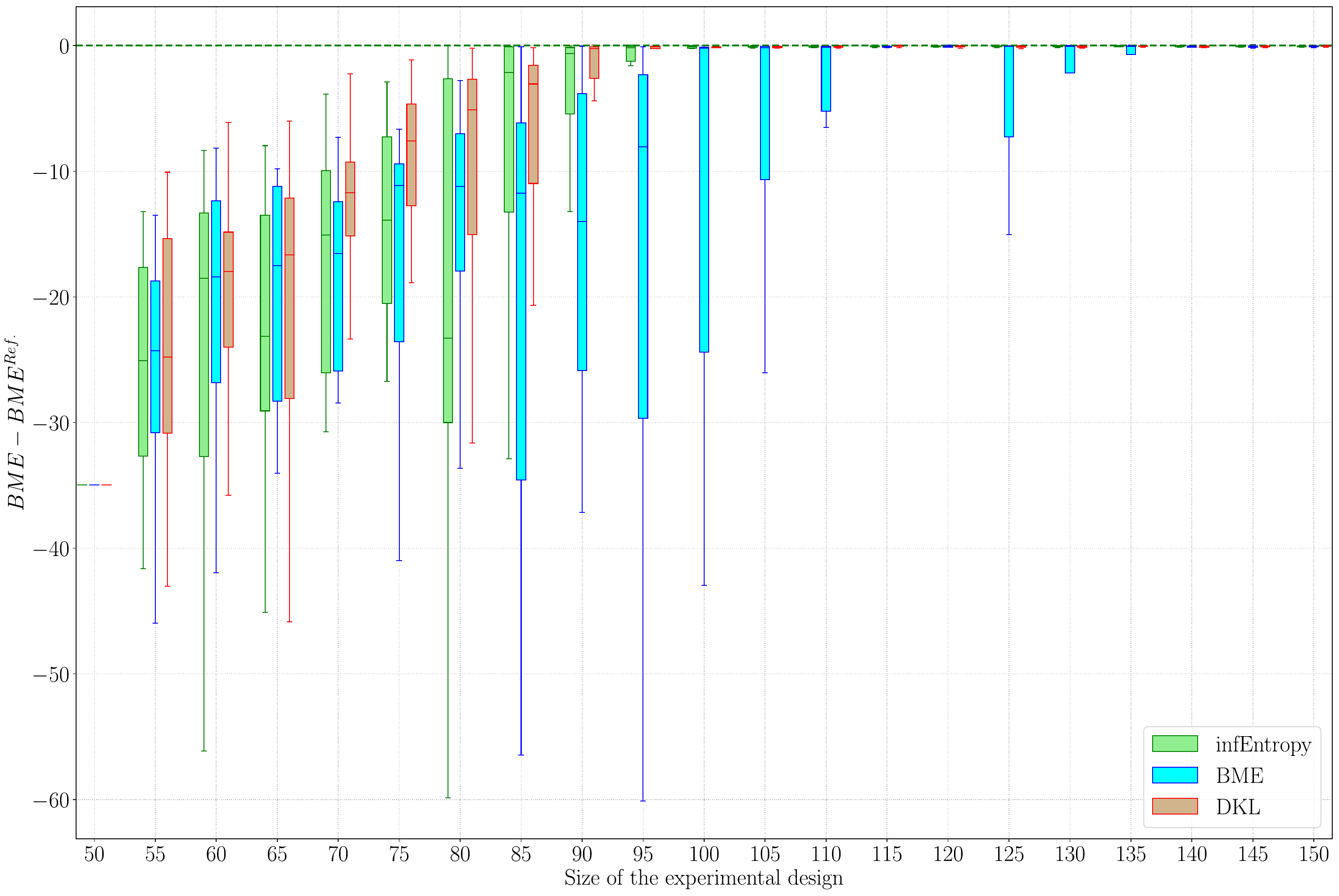}  
  \caption{Bayesian model evidence}
  \label{fig:seqDesignNum_BME}
\end{subfigure}
\begin{subfigure}{1.0\textwidth}
  \centering
  \includegraphics[width=1.0\linewidth]{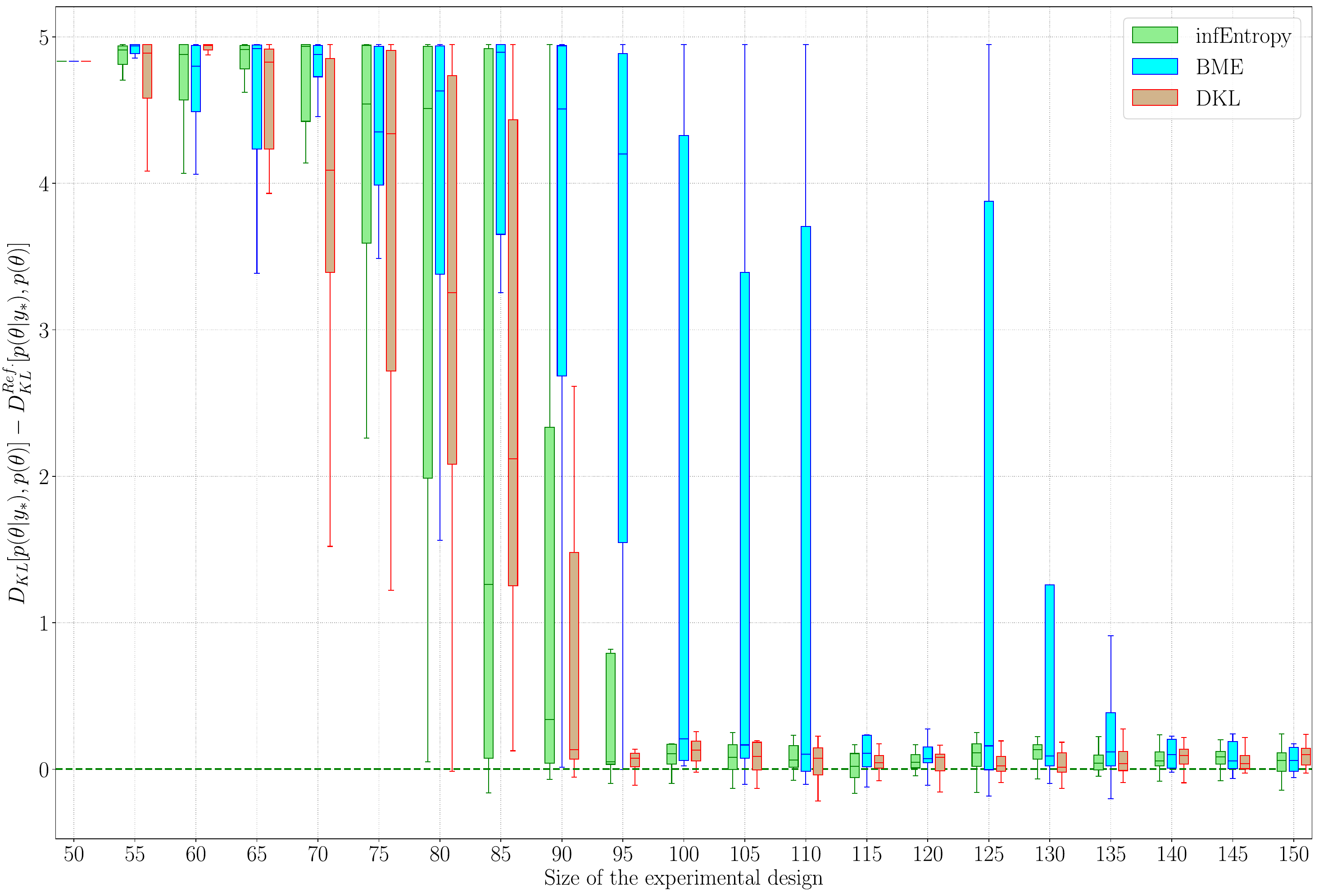}  
  \caption{Kullback-Leibler divergence}
  \label{fig:seqDesignNum_DKL}
\end{subfigure}
\caption{The convergence test of three Bayesian optimal design approaches with (a) Bayesian model evidence and (b) Kullback-Leibler divergence}
\label{fig:seqDesignNum}
\end{figure}

Figure \ref{fig:seqDesignNum} reveals that with only 150 runs for a highly non-linear problem with the parameter space of 10 dimensions, the BME and DKL values converge to the reference value, obtained by running a million of model evaluations. It can also be noticed that the information gain utility (DKL) and information entropy based utility (infoEntropy) have reached an acceptable convergence after only 100 runs.

%% file: TEXs/Application.tex
\chapter{Application to Flow Simulation Models for Fractured Porous Media}
\thispagestyle{empty}

Flow in porous media is often characterized by very strong heterogeneities,in particular fractures,  whose influence is important for the understanding of the overall systems’ behavior in many natural and technical applications. There  are  many  competing model concepts to represent the flow in fractured porous media. These models vary drastically in their level of geometric details, simplifications and abstraction. In this chapter, we apply the Bayesian Validation Framework, introduced earlier, to set up a benchmark for two models of flow simulation in fractured porous media. 

\section{Experimental setup}
\begin{figure}[h!]
\begin{subfigure}{.475\textwidth}
  \centering
  \includegraphics[width=.925\linewidth]{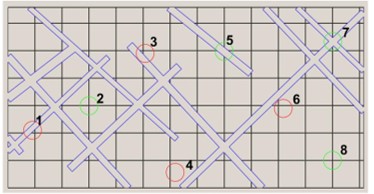}  
  \caption{Setup with connected fractures}
  \label{fig:ExpSetup_con}
\end{subfigure}
\begin{subfigure}{.475\textwidth}
  \centering
  \includegraphics[width=.95\linewidth]{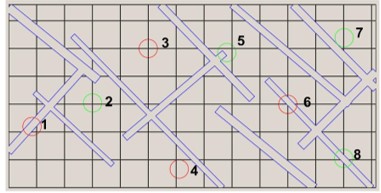}  
  \caption{Setup with disconnected fractures}
  \label{fig:ExpSetup_discon}
\end{subfigure}
\caption{Two experimental setups: (a) connected and (b) disconnected fracture network}
\label{fig:ExpSetup}
\end{figure}

Two different cases were considered in the experiments. In the first case, the fractures are connected from one side of the sample to the other (Figure \ref{fig:ExpSetup_con}), while they are disconnected in the second case (Figure \ref{fig:ExpSetup_discon}).
In addition, measuring pressure values at in-/outlet and eight intermediate points of disconnected/connected fracture network sample were accomplished with six tested and calibrated sensors (four at the intermediate points and two at in-/outlet locations).

Due to lack of space on the sample for installation of sensor at all 8 measurement points, the pressure values were measured separately on the sample, four sensors at the time, for different flow rate. 
The fluid used in this experiment was distilled water and flow-rate of the fluid was carefully adopted in range of 0.1 to 0.5 [ml/min] not to induce critical deformation into the sample which may cause pressure diffusion.

\section{Forward models}
Two model variants for flow simulation in fractured porous media have been investigated in this study. The first model (B01) employs a phase-field fraction representation, while the second (B03) makes use of a sharp fracture model.
The model B01 simulates the solid fluid interaction within the porous medium under the consideration of Biot’s Theory of consolidation. Moreover, the flow of the fluid in the porous medium is described by Darcy’s law where the permeability within the cracks is increased by an additional permeability tensor. It models Poiseuille-type flow within the cracks and is derived by the lubrication theory. 
The solution is obtained by the finite element method in \textit{FEAP}\cite{taylor_feap_2003}. 

The Model B03 follows a discrete fracture network approach. All fractures are geometrically resolved conforming to the mesh.
A mixed-dimensional model is used for Darcy flow both in the bulk porous medium and the fractures. The permeability of the fractures is determined by the Poiseuille approximation using the hydraulic diameter.
The solution is obtained by a Finite-Volume method implemented in \textit{DuMu$^x$}\cite{Kochetal2020Dumux}.

\begin{figure}[h!]
\begin{subfigure}{.475\textwidth}
  \centering
  \includegraphics[width=.925\linewidth]{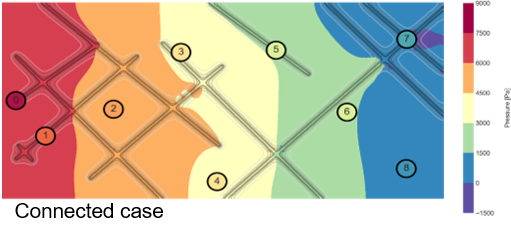}  
  \caption{Model B01}
  \label{fig:modelB01}
\end{subfigure}
\begin{subfigure}{.475\textwidth}
  \centering
  \includegraphics[width=.95\linewidth]{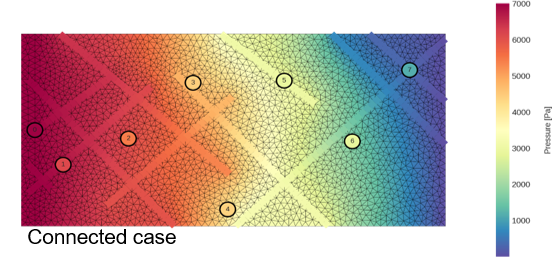}  
  \caption{Model B03}
  \label{fig:modelB03}
\end{subfigure}
\caption{The model domains of the connected case for (a) Model B01 and (b) Model B03}
\label{fig:models}
\end{figure}

Uncertain parameters and their ranges for the model validation analysis are summarized in Table \ref{Tab:Params_Ranges}.
\begin{table}[h!]
\caption{The list of considered uncertain parameters and their associated distribution as prior knowledge}
\centering
\begin{tabular}{ |c|c|c|c| }
\hline
Parameter name & Range & Distribution type \\
\hline \hline
Permeability (porous medium) & $\left[10^{-13},10^{-11}\right]$ & uniform \\ 
Permeabiliy (fractures) & $\left[10^{-8},10^{-7}\right]$ & uniform \\ 
Sample's depth & $\left[8 \times 10^{-5},10^{-4}\right]$ & uniform \\ 
\hline
\end{tabular}
\label{Tab:Params_Ranges}
\end{table}

\section{Error quantification}
Bayesian statistics updates the prior belief on the model by comparing the model responses with the measured data. It can include not only the errors and uncertainties in the observed data, but it can also take into consideration other sources of errors, such as numerical errors and surrogate model's error. These errors can be considered in the covariance matrix $\Sigma$ in Eq. \eqref{eq:BayesLikelihood}, assuming that they are follow a normal distribution.  
In the following, we will shed light on three sources of error used in the validation benchmark of the flow simulation models for fractured porous media.

\subsection{Measurement errors}
\begin{figure}[h!]
\begin{subfigure}{.475\textwidth}
  \centering
  \includegraphics[width=.95\linewidth]{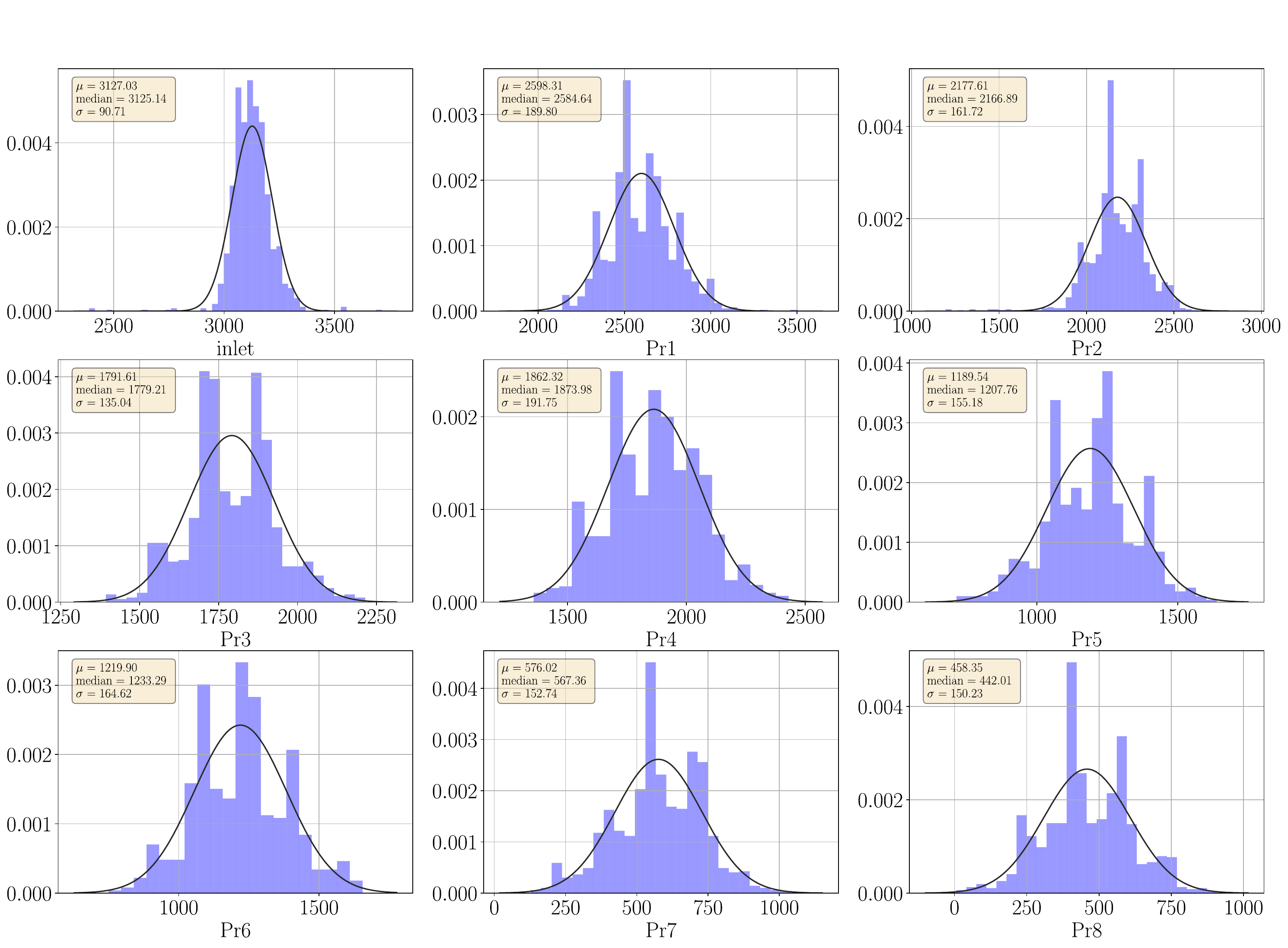}  
  \caption{Calibration with $Q=0.2 \: ml/min$}
  \label{fig:ExpError_con}
\end{subfigure}
\begin{subfigure}{.475\textwidth}
  \centering
  \includegraphics[width=.95\linewidth]{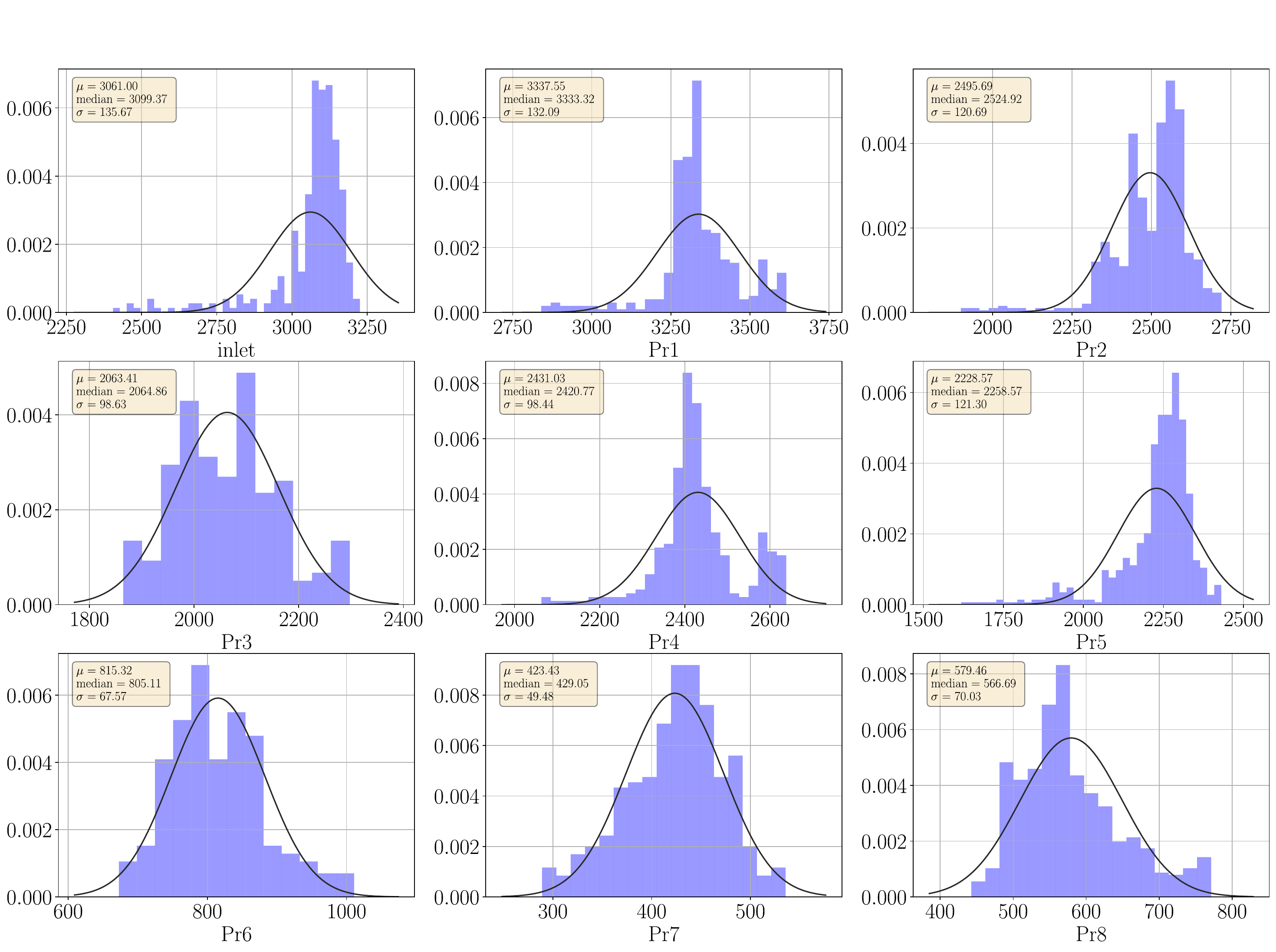}  
  \caption{Calibration with $Q=0.1 \: ml/min$}
  \label{fig:ExpError_discon}
\end{subfigure}
\caption{The pressure distribution for 9 sensors for (a) the connected and (b) the disconnected case}
\label{fig:ExpError}
\end{figure}

We use the standard deviation of the measured value for each measurement sensor as the experimental error. These values are extracted for two cases and for different flow rates for the calibration and validation, separately. The pressure measurement's distribution for two different cases for the calibration step (\textit{i.e.} with the flow rate 0.2 [ml/min]) are shown in Figure \ref{fig:ExpError}. 

\subsection{Numerical errors}
Here, we only investigate the discretization error that originates from a certain choice of the meshing size. Following \cite{oberkampf2010verification}, we take a heuristic approach to quantify this error is to make a comparison between different mesh spacing and to make use of a generalized Richardson extrapolation to estimate the error. The Richardson extrapolation takes the following form:
\begin{equation}
    \label{eq:RichardsonExtrapol}
    f_{k}=\bar{f}+g_{p} h_{k}^{\hat{p}}+\mathcal{O}\left(h_{k}^{\hat{p}+1}\right)
\end{equation}
$f_{k}$ denotes the exact solution to the discrete equation on a mesh with spacing $h_{k}$ (known), $\bar{f}$ stands for the exact solution to the original PDE (unknown). $g_{p}$ is the error term coefficient and $\hat{p}$ indicates the observed order of accuracy. Here, we seek the error with an order of one. Thus, the unknowns, \textit{i.e.} $\bar{f}$ and $g_{p}$, can be easily determined via a least square method. 

\subsection{Surrogate model's errors}
As discussed earlier, we replace the computationally intensive model in the Bayesian framework to offset the computational cost. By doing so, a new source of error is introduced, namely surrogate modeling's error. In this study, we set the limit for computational budget to 200 simulation runs for each model. The prediction error of the surrogate model, introduced in Section \ref{PredictBaSPCE}, is taken into accounts in the calculation of the likelihoods in Eq. \eqref{eq:BayesLikelihood}. 

\section{Bayesian model validation}
For the validation within the Bayesian approach, the following steps are taken. First, a surrogate model is trained based on the simulation results, obtained by the original computational models, here the pressure readings at 9 sensors shown in Figure \ref{fig:ExpSetup}.
Then, the Bayesian updating is performed in that the prior knowledge on the distribution of uncertain parameters is revised by comparing the model outputs with the measurement, using Eq. \eqref{eq:BayesInference}. This can be done either by a brute-force Monte Carlo sampling (\textit{a.k.a} rejection sampling) or a more sophisticated methods, for instance Markov-Chain Monte Carlo. Afterwards, the resulting posterior distribution is used to compare the model with a new observed data set. In what follows, some preliminary results will be presented.

\subsection{Preliminary results}
As discussed in Section \ref{BayesianModelComparison}, Bayesian model evidence (BME) provides a reasonable metric for assessment of the model performance. BME is sometimes also referred to as marginal likelihood because it quantifies the likelihood of the model to have produced the observed data averaged over the complete prior parameter space.
To also consider the uncertainty associated with this value and its influence on our analysis, the measured data has been perturbed with a Gaussian noise. 

Moreover, we use the distribution of BME for TOM, following \cite{schoniger2014model}, computed by a chi-square distribution, as upper limit for model performance, as defined in Section \ref{TOM}.
\begin{figure}[h!]
    \centering
    \includegraphics[width=.75\linewidth]{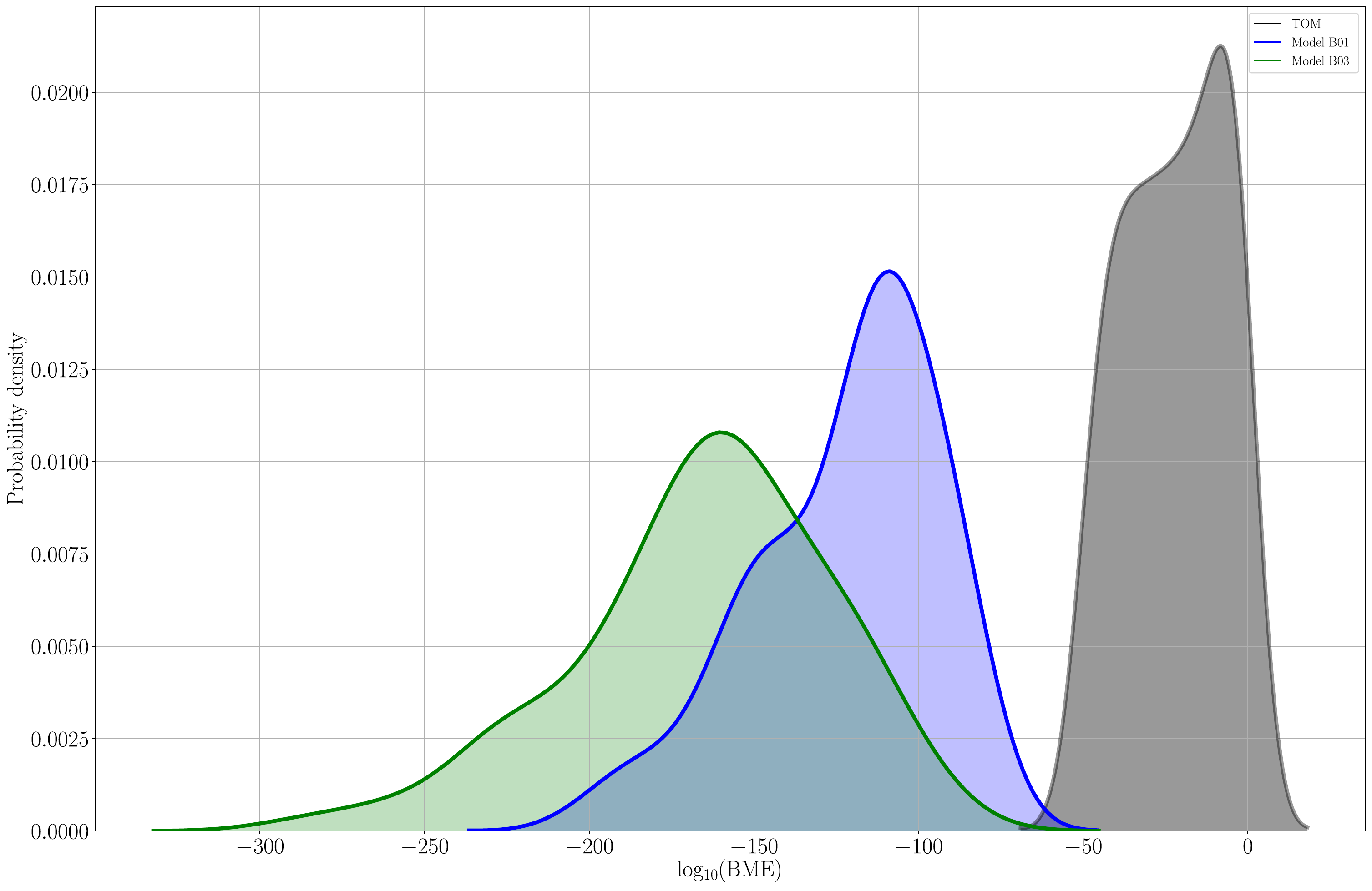}
    \caption{The BME distributions for the competing models with the perturbed data set}
    \label{fig:BME_TOM_Valid}
\end{figure}
Figure \ref{fig:BME_TOM_Valid} shows the distribution of the BME for the models B01 and B03, as well as the upper limit performance.These values are obtained by varying the measurement data set to account for the uncertainty associated with BME.
Model B01 shows slightly better performance, as its BME distribution is closer to that of TOM.

One approach to further compare models is hypothesis testing in a Bayesian setting. This can be done by a so-called Bayes factor, that has been presented in Section \ref{BayesHypoTest}.
The Bayes factor is a measure for significance in Bayesian hypothesis testing. It quantiﬁes the evidence of hypothesis that $M_l$ is the data generating model against a null-hypothesis.
The null-hypothesis can be defined as one model to be the best model among the set of models under investigation.
\begin{figure}[h!]
\begin{subfigure}{.475\textwidth}
  \centering
  \includegraphics[width=.95\linewidth]{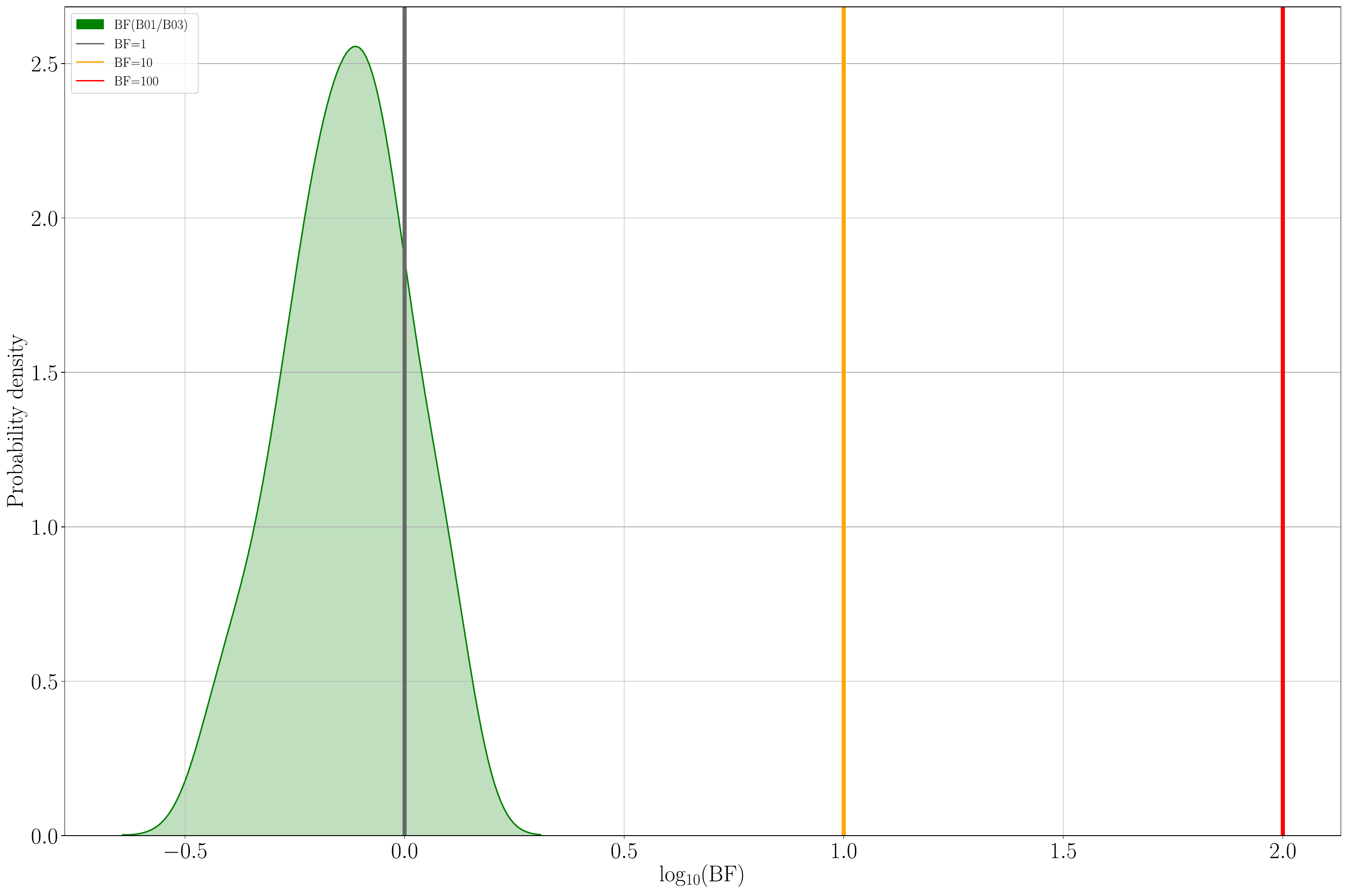}  
  \caption{Model B01 vs. model B03}
  \label{fig:BayesFactor_B01_B03}
\end{subfigure}
\begin{subfigure}{.475\textwidth}
  \centering
  \includegraphics[width=.95\linewidth]{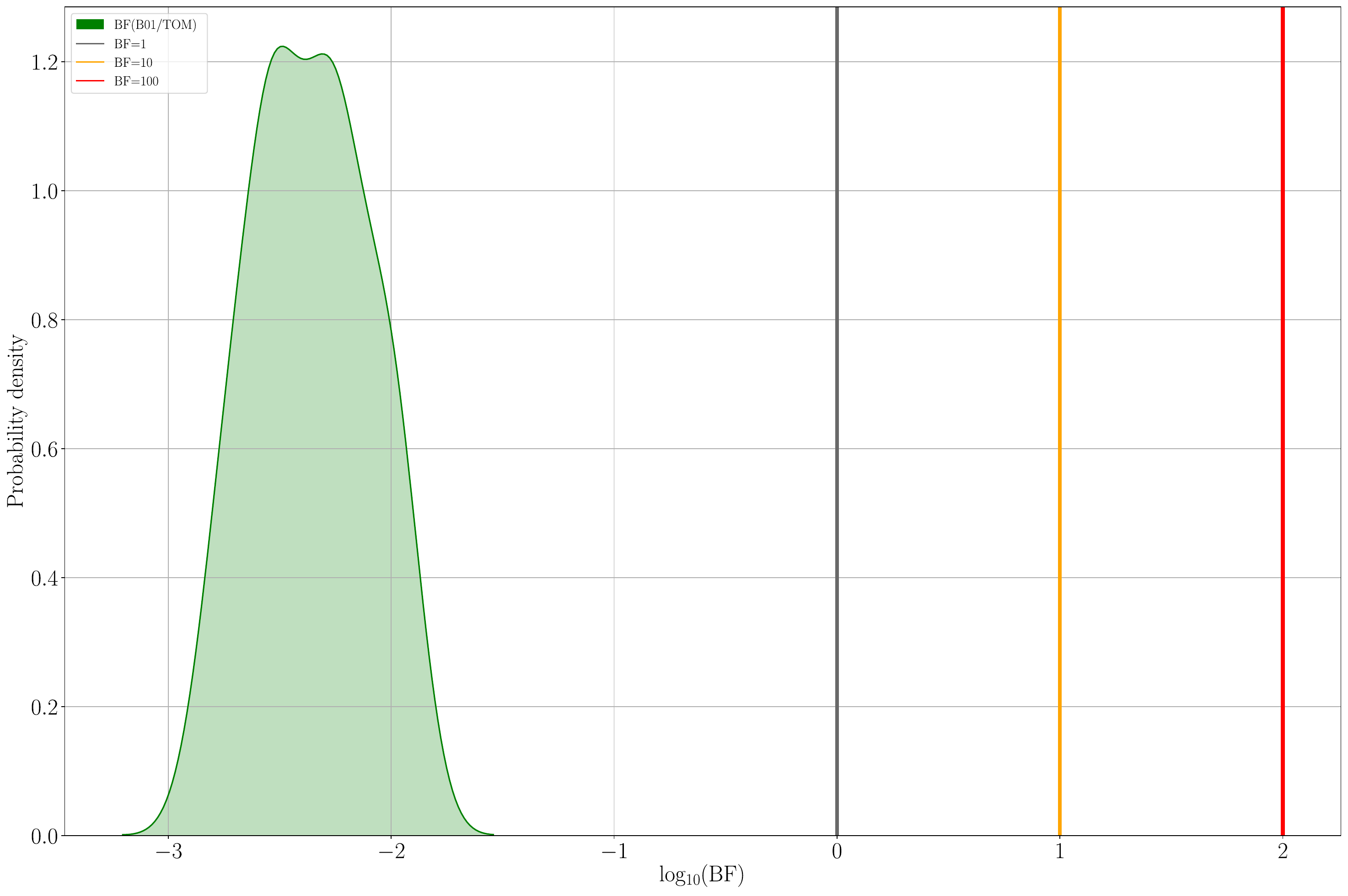}  
  \caption{Model B01 vs. TOM model}
  \label{fig:BayesFactor_B01_TOM}
\end{subfigure}
\caption{The Bayesian factor for two scenarios: (a) Model B03 performs better as the data-generating model (b) theoretically optimal model has the best performance}
\label{fig:BayesFactor}
\end{figure}
Figure \ref{fig:BayesFactor} illustrates the distribution of Bayes factor of two models in log-scale. 
The significance levels, suggested by Jeffreys \cite{jeffreys1961theory} are also marked here with vertical solid lines. One can interpret that there is no substantial evidence to favor B01 against B03, as shown in Figure \ref{fig:BayesFactor_B01_B03}.
We can also compare the performance of each model with the upper limit for performance, provided by the theoretically optimal model, in short “TOM”. In Figure \ref{fig:BayesFactor_B01_TOM}, the distribution of the Bayes factor includes negative values, that means the evidence is in favor of the null hypothesis instead of the actual hypothesis. Thus, it can be concluded that there is substantial evidence in favor of the TOM against Model B01.